%% file: main.tex
\documentclass[acmsmall]{acmart}

\AtBeginDocument{%
  }

\usepackage{booktabs}
\usepackage{multirow}
\usepackage{svg}
\usepackage{siunitx}
\usepackage{algorithm}
\usepackage{algpseudocode}
\usepackage{graphicx} 
\usepackage{pgfplots}
\usepackage[most]{tcolorbox}
\usepackage{dashrule}

\usepackage{enumitem}
\usepackage{listings}
\usepackage{xcolor}
\usepackage{float}
\usepackage{placeins}
\usepackage{tikz}
\usepackage{xspace}
\usepackage{adjustbox}
\usepackage{subcaption}

\definecolor{codegreen}{rgb}{0,0.6,0}
\definecolor{codegray}{rgb}{0.5,0.5,0.5}
\definecolor{codepurple}{rgb}{0.58,0,0.82}
\definecolor{backcolour}{rgb}{0.95,0.95,0.92}
\definecolor{errorred}{rgb}{0.6,0,0}

\lstdefinestyle{pythonstyle}{
    backgroundcolor=\color{backcolour},
    commentstyle=\color{codegreen},
    keywordstyle=\color{magenta},
    numberstyle=\tiny\color{codegray},
    stringstyle=\color{codepurple},
    basicstyle=\ttfamily\footnotesize,
    breakatwhitespace=false,
    breaklines=true,
    captionpos=b,
    keepspaces=true,
    numbers=left,
    numbersep=5pt,
    showspaces=false,
    showstringspaces=false,
    showtabs=false,
    tabsize=2,
    language=Python
}

\lstdefinestyle{promptstyle}{
    backgroundcolor=\color{backcolour},         
    basicstyle=\ttfamily\footnotesize\color{black}, 
    commentstyle=\color{black},           
    keywordstyle=\color{black},           
    stringstyle=\color{black},             
    numberstyle=\tiny\color{black},        
    breaklines=true,
    numbers=left,
    numbersep=10pt,
    tabsize=4,
    showstringspaces=false,
    keepspaces=true,
}

\newtcolorbox{RQsummary}{
  colback=gray!6!white,
  colframe=gray!60!black,
  boxrule=0.5pt,
  arc=2mm,
  left=1mm,
  right=1mm,
  top=1mm,
  bottom=1mm,
  before skip=5pt,
  after skip=5pt,
}

\pgfplotsset{compat=1.18}

\setcopyright{cc}
\setcctype{by}
\acmDOI{10.1145/3797144}
\acmYear{2026}
\acmJournal{PACMSE}
\acmVolume{3}
\acmNumber{FSE}
\acmArticle{FSE017}
\acmMonth{7}
\acmSubmissionID{fse26maina-p306-p}
\received{2025-09-11}
\received[accepted]{2025-12-22}

\newcommand{\method}{\textsc{Hydra}\@\xspace}

\begin{document}

\title{Do Not Treat Code as Natural Language: Implications for Repository-Level Code Generation and Beyond}

\author{Minh Le-Anh}
\email{minhla4@fpt.com}
\orcid{0009-0000-2859-7038}
\affiliation{%
  \institution{FPT Software AI Center and Hanoi University of Science and Technology}
  \city{Hanoi}
  \country{Vietnam}
}

\author{Huyen Nguyen}
\email{huyen.nt235507@sis.hust.edu.vn}
\orcid{0009-0009-4724-9050}
\affiliation{%
  \institution{Hanoi University of Science and Technology}
  \city{Hanoi}
  \country{Vietnam}}

\author{An Khanh Tran}
\email{khanh.ta225447@sis.hust.edu.vn}
\orcid{0009-0001-4175-7347}
\affiliation{%
  \institution{Hanoi University of Science and Technology}
  \city{Hanoi}
  \country{Vietnam}}

\author{Nam Le Hai}
\authornote{Nam Le Hai is the corresponding author: \href{mailto:namlh@soict.hust.edu.vn}{namlh@soict.hust.edu.vn}}
\email{namlh@soict.hust.edu.vn}
\orcid{0009-0005-8895-6051}
\affiliation{%
  \institution{Hanoi University of Science and Technology}
  \city{Hanoi}
  \country{Vietnam}}

\author{Linh Ngo Van}
\email{linhnv@soict.hust.edu.vn}
\orcid{0000-0002-0011-5137}
\affiliation{%
  \institution{Hanoi University of Science and Technology}
  \city{Hanoi}
  \country{Vietnam}}

\author{Nghi D.Q. Bui}
\email{bdqnghi@gmail.com}
\orcid{0000-0003-1984-4329}
\affiliation{%
  \institution{FPT Software AI Center}
  \city{Hanoi}
  \country{Vietnam}}

\author{Bach Le}
\email{bach.le@unimelb.edu.au}
\orcid{0000-0001-5044-1582}
\affiliation{%
  \institution{The University of Melbourne}
  \city{Melbourne}
  \state{Victoria}
  \country{Australia}}


\begin{abstract}
Large language models for code (CodeLLMs) have demonstrated remarkable success in standalone code completion and generation, sometimes even surpassing human performance, yet their effectiveness diminishes in repository-level settings where cross-file dependencies and structural context are essential. Existing Retrieval-Augmented Generation (RAG) approaches often borrow strategies from NLP, relying on chunking-based indexing and similarity-based retrieval. Chunking results in the loss of coherence between code units and overlooks structural relationships, while similarity-driven methods frequently miss functionally relevant dependencies such as helper functions, classes, or global variables. To address these limitations, we present \method, a repository-level code generation framework that treats code as structured code rather than natural language. Our approach introduces (i) a structure-aware indexing strategy that represents repositories as hierarchical trees of functions, classes, and variables, preserving code structure and dependencies, (ii) a lightweight dependency-aware retriever (DAR) that explicitly identifies and retrieves the true dependencies required by a target function, and (iii) a hybrid retrieval mechanism that combines DAR with similarity-based retrieval to provide both essential building blocks and practical usage examples. Extensive experiments on the challenging DevEval and RepoExec benchmarks, both requiring function implementation from real-world repositories with complex large repository context, show that \method achieves state-of-the-art performance across open- and closed-source CodeLLMs. Notably, our method establishes a new state of the art in repository-level code generation, surpassing strongest baseline by over 5\% in Pass@1 and even enabling smaller models to match or exceed the performance of much larger ones that rely on existing retrievers.

\end{abstract}

\ccsdesc[500]{Software and its engineering~Automatic programming}
\ccsdesc[500]{Information systems~Information retrieval}
\ccsdesc[500]{Computing methodologies~Artificial intelligence}

\keywords{Repository-Level Code Generation, Code Generation, Retrieval-Augmented Generation, Large Language Models}


\maketitle

\input{section/1_introduction}
\input{section/2_background}
\input{section/3_approach}

\input{section/4_experiment}

\input{section/5_evaluation}
\input{section/7.Threads}
\input{section/8_conclusion}
\input{section/9.data-avai}
\section*{Acknowledgments}

Nam Le Hai was funded by the PhD Scholarship Programme of Vingroup Innovation Foundation (VINIF), VinUniversity, code VINIF.2025.TS68.

\newpage
\bibliographystyle{ACM-Reference-Format}
\bibliography{references}
\end{document}

%% file: section/1_introduction.tex
\section{Introduction}
\label{sec:intro}

Large language models for code (CodeLLMs) have achieved impressive results in code completion and generation, but they typically operate within a single file or limited context \citep{austin2021mbpp, chen2021evaluatinglargelanguagemodels, chencodet, le2022coderl, wang2021codet5, nguyen2025codemmlu}. In real-world software development, completing code often requires incorporating repository-level context or cross-file context \citep{nam2024repoexec, Repoformer, zhang2023repocoderrepositorylevelcodecompletion, DevEval, deng2024r2c2coder, jain2024r2e}. Retrieval-Augmented Generation (RAG) has emerged as a promising approach to address this need \citep{cao2024retrievalaccurategeneration, zhang2023repocoderrepositorylevelcodecompletion, Repoformer, wang2025coderag}. In repository-level code completion, a retrieval module fetches relevant code snippets from across the codebase to supply the model with cross-file information beyond the current file. This paradigm has shown strong empirical gains, bridging the gap in cross-file knowledge. Recent work has further improved RAG by designing better retrieval mechanisms and structured prompts for code LMs \cite{Repoformer}. However, most existing approaches treat code as if it were natural language, directly borrowing techniques from the NLP domain. This leads to several key challenges that we aim to overcome.

\paragraph{\textbf{Limitations of Chunking-Based Indexing.}} Multiple prior code generation RAG-based systems index a codebase by breaking files into sequential chunks of code text \citep{Repoformer, zhang2023repocoderrepositorylevelcodecompletion, wang2024rlcoderreinforcementlearningrepositorylevel, liang2024repofuse}. For example, files are split into fixed-size snippets (e.g. 20-line blocks) with sliding windows , which are then used to build a retrieval index. At query time, the model’s context is matched against these chunks using similarity search. While this chunking strategy is common in long-document NLP tasks \citep{sarthi2024raptor, edge2024local, gao2024retrievalaugmentedgenerationlargelanguage}, it is poorly aligned with the structural nature of code: it fragments logically coherent units (e.g. splitting a function or class), includes irrelevant surrounding code context, and assumes that relevant context must be contiguous in text. In reality, the most important information for code completion often lies in structure code blocks such as function definitions, class implementations, or imported modules that may reside in entirely different files. Unfortunately, most existing methods for repo-level code generation that often chunk files into continuous blocks \citep{wang2024rlcoderreinforcementlearningrepositorylevel, Repoformer, zhang2023repocoderrepositorylevelcodecompletion} fail to retrieve context that fully reflects this behavior. Treating code as flat text chunks ignores its hierarchical and dependency-driven structure, ultimately limiting retrieval effectiveness in repository-level tasks.

\paragraph{\textbf{Limitations of Similarity-Based Retrieval.}} Similarity-driven methods such as BM25, TF-IDF, Jaccard similarity, or embedding-based cosine search have been widely applied in document and  code search \citep{da2020crokage, husain2019codesearchnet, nguyen2023vault, li2024coir, guo2022unixcoderunifiedcrossmodalpretraining}, where the goal is to retrieve a semantically or lexically relevant snippet given a natural language or code query. In that setting, ranking by surface similarity often suffices, since the user primarily seeks related examples or references. However, in repository-level code generation, the requirements are fundamentally different. Generated code must align with the current state of the repository, respect existing dependencies, and follow the coding conventions of the project. Pure similarity matching is suboptimal for this, as functionally relevant context (e.g., a helper function definition, a data structure declaration, or an API implementation) may not share obvious lexical overlap with the target code. As a result, models relying only on similarity retrieval can miss crucial dependencies or introduce superficially related but irrelevant code. Recent studies \citep{nam2024repoexec, jain2024r2e} underscore that: providing a model with the ground-truth dependency context (all functions/classes that the code calls) yields significant gains in repository-level generation, whereas missing these dependencies leads to errors or redundant reimplementation. \citet{gu2025retrieve} further show that similarity-based retrieval ``struggle to capture code semantics'' and may even degrade generation quality by introducing noisy context, sometimes reducing accuracy. These findings underscore that while similarity retrieval is effective for code search, it might be insufficient for the task of code generation in repository-level settings.

To bridge these gaps, we propose \method, a repository-level code generation framework that explicitly exploits the structure and dependencies in a codebase. Specifically, \method is designed to retrieve and utilize repository context in a way that mirrors how developers themselves would navigate code, addressing the shortcomings identified above. Our contributions are summarized as follows:

\begin{enumerate}[label=\textbf{\raisebox{0.3pt}{\Large\textcircled{\small\arabic*}}}, leftmargin=2em]

    \item We present a method, Structure-aware indexing, to represent the entire repository as a hierarchical tree of code components (such as functions, classes, and global variables). Rather than splitting files into arbitrary text chunks, we parse and index the code at the level of logical units. This structured representation preserves critical relationships (for example, which functions belong to a class, or which files import which modules) that would be overlooked if flat text chunking is applied.


    \item We introduce a retrieval mechanism centered on \textit{call-graph dependencies}, namely dependency-aware retriever (DAR). Given a code generation query, the DAR automatically identifies referenced symbols and retrieves their definitions or implementations from the repository. To make this practical, we also propose a lightweight model with a fine-tuning strategy that empowers efficient and accurate dependency retrieval without heavy computational overhead. By explicitly modeling dependencies rather than relying on similarity alone, DAR fills a critical gap left by prior work.

    \item We propose \method Retriever, a hybrid retrieval strategy that combines DAR with similarity-based retrieval (BM25). While DAR ensures that the model has access to the correct building blocks (functions, classes, variables), similarity-based retrieval complements this by providing usage examples of how these dependencies are invoked in practice. This design yields a richer and more reliable context, substantially improving repository-level code generation.

    \item We conduct extensive experiments on two recent repository-level benchmarks, DevEval \citep{DevEval} and RepoExec \citep{nam2024repoexec}, focusing on function-level code generation in Python. Using both open-source (Qwen2.5-Coder, 1.5B \& 7B) and closed-source (GPT-4.1 mini) models, we show that \method generalizes across model families and sizes. It achieves state-of-the-art results, surpassing the strongest baselines with close-sourced generator by more than 5\%, and notably allows a 1.5B model to match or surpass a 7B model with existing retriever or no retrieval, effectively bridging a $\sim4\times$ size gap.

\end{enumerate}

%% file: section/2_background.tex
\section{Background \& Related Work}

\subsection{Large Language Models for Code}
\label{sec:codellms}
The rise of large language models pretrained on code (CodeLLMs) has transformed code generation. Open-source like StarCoder \citep{lozhkov2024starcoder} and CodeLlama \citep{roziere2023code} provided multilingual support and achieved state-of-the-art performance on benchmarks like MultiPL-E~\citep{cassano2023multipl}. More recently, instruction-tuned models such as DeepSeek-Coder\citep{guo2024deepseek} and Qwen-Coder\citep{hui2024qwen2} have improved controllability and debugging, narrowing the gap with proprietary models. Closed-source models have pushed the frontier further. OpenAI Codex \citep{chen2021evaluatinglargelanguagemodels} pioneered zero-shot code generation, later extended by GPT-4/4o \citep{achiam2023gpt, hurst2024gpt} with code interpreter functionalities. Anthropic’s Claude 3 and Google’s Gemini ~\citep{team2023gemini, team2024gemini} also demonstrate strong reasoning and multilingual coding ability, with enhanced support for repository-level understanding and integration into developer workflows.

\subsection{Retrieval-Augmented Generation}
\label{sec:rag}

Retrieval-Augmented Generation (RAG) \cite{gao2024retrievalaugmentedgenerationlargelanguage, zhao2024retrievalaugmentedgenerationaigeneratedcontent, sarthi2024raptor} is a general framework that enhances generative models by providing them with external, task-relevant context. Instead of relying solely on a model’s parametric knowledge, RAG augments the input with information retrieved from a supporting corpus. For repository-level code generation, this means treating the project’s entire codebase as the external knowledge source \citep{zhang2023repocoderrepositorylevelcodecompletion, Repoformer, li2024repomincoder}, enabling the model to access cross-file context beyond the current file.

Conceptually, RAG is composed of two modules: a retriever and a generator. The retriever is responsible for locating relevant pieces of information given a query, which in this domain may be an incomplete function or surrounding code context. It searches the codebase for snippets that are most likely useful to complete the query. These retrieved elements are then combined with the original prompt to form an enriched context. The generator, typically a large code model (Section \ref{sec:codellms}), consumes this augmented input and produces the final output, such as completing a function or generating a new implementation. We can formalize this interaction for a code generation task. Given a prompt \( P \) and a knowledge corpus \( K \) representing the target repository, the final generated code, \( C_{\text{final}} \), is produced by the function:
\begin{equation}
C_{\text{final}} = \mathcal{G}(P \oplus \mathcal{R}(P, K))
\end{equation}
where \( \mathcal{R}(\cdot) \) is the retrieval function that returns a set of relevant code chunks or documents from \( K \). The operator \( \oplus \) denotes the augmentation (e.g. concatenate) process, where the retrieved information is integrated with the original prompt \( P \). The generator function \( \mathcal{G}(\cdot) \) then synthesizes the final code based on this augmented input.

In most prior work, the retrieval component \( \mathcal{R}(\cdot) \) has been approached through either sparse syntactic methods, such as BM25, which rely on lexical overlap, or dense semantic methods, which leverage embedding similarity to capture semantic relatedness.

\begin{enumerate}
    \item \textbf{Syntactic sparse retrieval using BM25~\cite{article}}, which relies on lexical overlap between the query \( q \) and the code snippet \( d \).
    \begin{equation}
    \label{eq:bm25}
    \text{BM25}(q, d) = \sum_{t \in q} \text{IDF}(t) \cdot \frac{f(t, d) \cdot (k_1 + 1)}{f(t, d) + k_1 \cdot \left(1 - b + b \cdot \frac{|d|}{\text{avgdl}}\right)}
    \end{equation}
    where:
    \begin{itemize}
        \item $k_{1}$ is a saturation parameter controlling the impact of term frequency (typically $1.2 \leq k_{1} \leq 2.0$), $b$ is the length-normalization parameter ($0 \leq b \leq 1$),
        \item $avgdl$ is the average document length across the collection,
    \end{itemize}
    \item \textbf{Semantic dense retrieval using embedding similarity} is typically implemented with a pretrained encoder (e.g., UniXCoder~\citep{guo2022unixcoderunifiedcrossmodalpretraining} or CodeBERT~\citep{feng2020codebert}), which maps both the query $q$ and candidate document $d$ into vector representations. A similarity score (commonly cosine similarity) is then computed between these vectors to rank and retrieve the most relevant snippets.
    \begin{equation}
    \text{cosine\_sim}(q, d) = \frac{q \cdot d}{\|q\| \cdot \|d\|}
    \end{equation}
    where $\| \cdot \|$ denotes the L2 norm (also called the Euclidean norm) of a vector.
 
\end{enumerate}

\subsection{Repository-Level Code Generation}\label{sec:related_work}
Recent advances in Large Language Models (LLMs) have demonstrated remarkable progress in standalone code generation, where the goal is to produce a self-contained function or snippet given a natural language description or partial code \citep{chen2021evaluatinglargelanguagemodels, austin2021mbpp, hendrycks2021apps, nguyen2023vault}. While impressive, such settings often oversimplify real-world software development, where code rarely exists in isolation. In practice, completing or generating code usually requires access to repository-level context-including functions, classes, modules, and APIs defined across multiple files. This has drawn increasing attention from both academia and industry to the task of repository-level code generation, which better reflects practical development scenarios and has led to the introduction of several dedicated benchmarks \citep{nam2024repoexec, jain2024r2e, DevEval, xie2024codebenchgen, ding2023crosscodeeval, ding2024cocomic, liurepobench}. Formally, the task can be viewed as producing the final code given a query and relevant information drawn from the entire repository. In this work, we restrict our study to function-level code generation in Python within a repository-level context, reflecting the practical design of most benchmarks \citep{DevEval, nam2024repoexec, zhang2023repocoderrepositorylevelcodecompletion, yu2024codereval}, where the query typically includes a function signature and, optionally, its docstring.

Previous methodologies have primarily adopted the RAG framework (Section \ref{sec:rag}) to incorporate cross-file context for tackling this task. For instance, RepoCoder~\cite{zhang2023repocoderrepositorylevelcodecompletion} employs an iterative RAG mechanism, where code generated in a previous step is used to enrich the context for the subsequent turn. RepoFormer~\cite{Repoformer} introduces a Self-selective RAG, which allows the model to trigger a special token to decide whether retrieving external context is necessary. Meanwhile, RLCoder~\cite{wang2024rlcoderreinforcementlearningrepositorylevel} utilizes reinforcement learning to train its retriever encoder, RLRetriever, aiming to close the vector space gap between a query and its most useful context. $\text{A}^3$-CodeGen~\cite{liao2023a3codgen} extracts, fuses, and feeds three types of repository information into the LLM: local-aware, global-aware, and third-party-library information. R2C2-Coder~\cite{deng2024r2c2coder} introduces R2C2-Enhance, a prompt assembling method which retrieves from the constructed candidate pool for each completion cursor position. Another method, GraphCoder~\cite{liu2024graphcoder}, first builds a code context graph which contains control flow, data, and control dependence between code statements. It then performs coarse-to-fine retrieval of context-similar code snippets using this graph. RepoMinCoder~\cite{li2024repomincoder} uses an additional round of screening and ranking based on information loss to complement the original RAG process. RepoFuse~\cite{liang2024repofuse} combines analogy context and rationale context, then compresses them into prompts with restricted size.

While these techniques have demonstrated promising results, they also highlight several persistent challenges in the field: non-adaptive chunking methods that lack sufficient semantic-level understanding of code structure, and naive similarity measures incapable of capturing the functional relationships among programming elements.

%% file: section/3_approach.tex
\section{Approach}
\subsection{Overview}

\begin{figure*}
    \centering
    \includegraphics[width=0.95\linewidth]{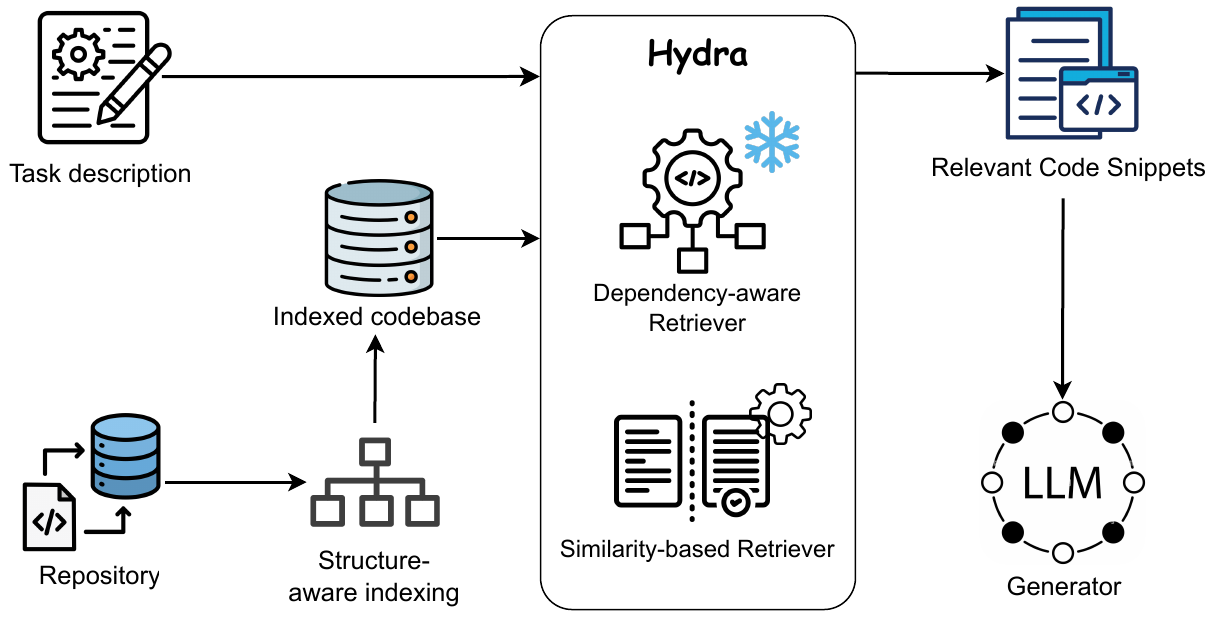}
    \caption{Overview of our workflow. The approach integrates structure-aware indexing, dependency-aware retrieval, and hybrid context construction to support repository-level code generation.}
    \label{fig:overview}
\end{figure*}

Figure \ref{fig:overview} provides an overview of the proposed approach, which operates as follows:

\begin{enumerate}[leftmargin=*]
    \item It begins with a task description (e.g., an incomplete function) and a repository as inputs. The repository is parsed into fine-grained components (e.g., classes, functions, variables) using an AST parser, which preserves both the overall tree structure of the repository and the complete implementation of each component. The codebase is then indexed using our structure-aware indexing approach.

    \item Next, retrieval is performed in two complementary ways: (i) the Dependency-aware retriever predicts and extracts the relevant dependencies for the target function by traversing the corresponding tree nodes, guided by the current file and its imported modules, and (ii) a similarity-based retriever (BM25) provides the top-k most similar snippets based on lexical similarity. These two sources are then merged to form a unified context that includes both dependency definitions and representative usage examples.

    \item Finally, the enriched context is concatenated with the task description and passed into a generator model (e.g., LLMs) to produce the completed code.
    
\end{enumerate}

\subsection{Structure-Aware Indexing}
\label{sec:indexing}

Unlike prior approaches that segment the codebase into arbitrary text chunks, our method preserves the semantic and structural integrity of the repository. Instead of viewing code as flat text, we treat it as a hierarchy of components derived from the program’s Abstract Syntax Tree (AST) using Python's \texttt{ast}\footnote{\url{https://docs.python.org/3/library/ast.html}} module. This allows us to represent the codebase as a graph of nodes, where each node corresponds to a fine-grained unit of code, and edges capture the relationships among them (e.g., imports or usage links). Specifically, we focus on three fundamental structures that are ubiquitous across programming languages:

\begin{itemize}
\item \textbf{Function nodes}, representing individual function or method implementations.

\item \textbf{Class nodes}, encapsulating class definitions and their member methods.

\item \textbf{Variable nodes}, including global or module-level variables that influence program behavior.
\end{itemize}

\begin{figure*}
    \centering
    \includegraphics[width=0.95\textwidth]{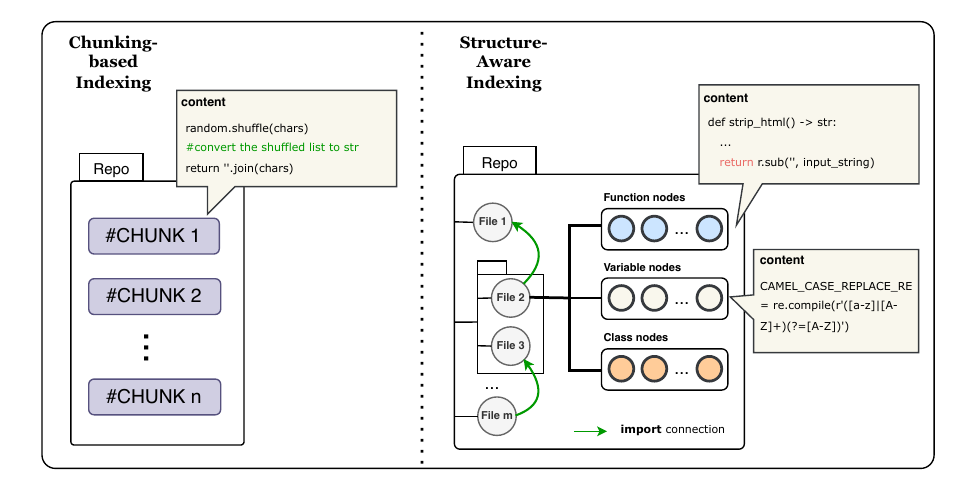}
    \caption{Comparison of chunk-based indexing and structure-aware indexing. Our approach preserves full functions, classes, and variables as nodes with structural links, avoiding the fragmentation and noise of traditional chunking.}
    \label{fig:indexing}
\end{figure*}

Each node preserves the complete implementation content of the corresponding component, rather than fragmenting it across multiple chunks. For example, a function node contains the entire body of the function, ensuring that retrieval always yields a self-contained and executable unit. Figure \ref{fig:indexing} contrasts our structure-aware indexing with conventional chunk-based indexing. In chunking, the repository is split into sequential blocks of code lines, which often fragment semantically related components (e.g., splitting a function across chunks or grouping unrelated code together). This fragmentation leads to noisy retrieval and loss of critical contextual information. By contrast, our approach organizes the repository into nodes aligned with the program’s natural structure. Dependencies across files (e.g., imports) are explicitly captured as edges between nodes, allowing retrieval methods to navigate not only by textual similarity but also by structural relationships.

This design enables more accurate and contextually faithful retrieval for repository-level code generation. It ensures that the model can access the precise functions, classes, or variables required to complete a task, while maintaining their full definitions and preserving the hierarchical organization of the codebase.

\subsection{Dependency-Aware Retrieval}

\paragraph{\textbf{Motivation.}} Previous methods commonly rely on NLP-inspired similarity-based retrieval \citep{article, devlin2019bert, chen2024bge, nguyen2025improving, truong2025emo, an2026mol}, such as BM25 or embedding models like UniXCoder, to support RAG for repository-level code generation. While these methods can surface code snippets that are lexically or semantically similar, they are poorly suited to capture the structural and functional relationships inherent in code. In practice, the most relevant context for generation is often not the ``most similar'' snippet, but rather the precise functions, classes, or variables that the target function depends on. Recent work, such as RepoExec~\citep{nam2024repoexec}, highlights that without access to ground-truth dependencies, models might produce low-quality code and accumulate technical debt \citep{li2023automatic,hai2024dopamin,russo2025leveraging, hai2025detectiontechnicaldebtjava}. However, RepoExec primarily establishes a benchmark and evaluation setting; it does not address the practical challenge of how to automatically retrieve these dependencies from a large codebase. Our empirical analysis confirms this gap: when relying solely on similarity-driven retrieval methods such as BM25 or UniXCoder, \textit{40–60\% of ground-truth dependencies required by the target function are missing from the retrieved context}. This shortfall demonstrates that similarity-based retrieval alone fails to expose the fine-grained dependencies essential for accurate repository-level code generation.

\paragraph{\textbf{Solution Approach.}} To address this challenge, we introduce a lightweight dependency-aware retriever (DAR) that is fine-tuned to detect relevant dependencies for a given query (i.e., a function signature and its accompanying docstring, if available). Training this retriever requires a dataset of triplets $(q, d^{+}, d^{-})$, where $q$ denotes the query, $d^{+}$ is a set of true dependencies, and $d^{-}$ is the set of irrelevant candidates. To construct such data, we mined Python repositories from GitHub and processed them using the \textit{structure-aware indexing} procedure described in Section~\ref{sec:indexing}. The full data preparation pipeline is detailed in Section~\ref{sec:training_data}. With this dataset, we fine-tune an encoder, framing dependency detection as a \textit{binary classification task} over query--candidate pairs. Specifically, the model is trained to predict whether a pair $(q, d)$ corresponds to a valid dependency, as described in Section~\ref{sec:classification}.
 
\pagebreak
\subsubsection{Training Dataset Construction} 
\label{sec:training_data}

\paragraph{\textbf{Data Source.}} To construct a high-quality dataset for training our retriever, we collected Python repositories from GitHub that met the following criteria: each repository had at least 30 forks, 10 stars, and a minimum of 5 Python files, resulting in 2,864 repositories. To avoid data contamination, we excluded any repositories that overlapped with the evaluation datasets. From this curated collection, we extracted all function, class, and variable blocks as in Section \ref{sec:indexing}, ensuring that the dataset preserved the structural granularity of source code.

\paragraph{\textbf{Triplets Construction.}} 
The training data for our retriever is organized into triplets $(q, d^{+}, d^{-})$. The detailed process is presented in Algorithm~\ref{alg:method_retriever_final_triplets}. To create the query $q$, we extract function nodes set $\mathcal{F}_{target}$ from the crawled repositories and use their function signature together with the docstring (if available) as the query. Since our objective is to retrieve the dependencies that the target function directly calls, we restrict candidates to components appearing in the same file or in its imported files. As illustrated in Figure~\ref{fig:indexing}, we traverse these files through the import connections to collect all components (functions, classes, and variables). For each anchor function $f_{i} \in \mathcal{F}_{target}$, components that are explicitly invoked by $f_{i}$ are added to the positive set $d_i^{+}$, while all others serve as negative samples $d_i^{-}$, to form $(q_i, d_i^+, d_i^-)$.

\begin{algorithm}[t]
\small
\caption{Training Data Construction}
\label{alg:method_retriever_final_triplets}
\begin{algorithmic}[1]
    \Require Filtered set of Python repositories $R_{filtered}$.
    \Ensure Training dataset $D$ with $(q, d^{+}, d^{-})$ triplets.
    
    \State $D \leftarrow \emptyset$
    \ForAll{repository $r \in R_{filtered}$}
        \State $\mathcal{F}, \mathcal{C}, \mathcal{V} \leftarrow \text{ExtractAllCodeUnits}(r)$ \Comment{Function, Class, and Variable Pools}
        
        \ForAll{function $f_i \in \mathcal{F}$}
            \State $files_{rel} \leftarrow \text{GetFileOf}(f_i) \cup \text{GetImportedFilesBy}(f_i)$
            \State $S_{context} \leftarrow \text{FilterUnitsByFiles}(\mathcal{F}, \mathcal{C}, \mathcal{V}, files_{rel})$
            
            \State $q_i \leftarrow \text{GetSignature}(f_i) + \text{GetDocstring}(f_i)$ \Comment{Create the query}
            \State $d^{+}_i \leftarrow \text{AnalyzeDependenciesAST}(f_i, S_{context})$ \Comment{Positive set}
            \State $d^{-}_i \leftarrow S_{context} \setminus (d^{+}_i \cup \{f_i\})$ \Comment{Negative set}
            
            \State Append $(q_i, d^{+}_i, d^{-}_i)$ to $D$
        \EndFor
    \EndFor
    \State \Return $D$
\end{algorithmic}
\end{algorithm}

\subsubsection{Retriever Fine-Tuning Strategy} 
\label{sec:classification}

Once the training dataset is constructed (Section~\ref{sec:training_data}), we fine-tune an encoder (UniXCoder \citep{guo2022unixcoderunifiedcrossmodalpretraining}) to detect the true dependencies $d^{+}$ for a given query $q$. We considered two alternative training strategies.  

\begin{enumerate}
    \item \textbf{Contrastive learning:} Following common practice in code search, one option is to optimize a contrastive loss, encouraging embeddings of $(q, d^{+})$ pairs to be closer than $(q, d^{-})$ pairs. At inference, retrieval is then performed via embedding similarity (e.g., cosine similarity). This approach is computationally efficient for large-scale retrieval. 

    \item \textbf{Pairwise classification:} Alternatively, we treat each pair $(q, d)$ with $d \in d^{+} \cup d^{-}$ as an independent input and train the encoder to classify concatenated input $p\oplus d$ into \{0,1\}, where $1$ denotes a valid dependency and $0$ otherwise.  
\end{enumerate} 

While the contrastive approach can offer speed advantages during inference, in our setting it proved unstable and failed to converge, as below discussion. We therefore adopt the pairwise classification strategy, which yields more reliable training dynamics and more accurate dependency detection. As a result, this process yielded a dataset of approximately $564,740$ data samples. The data is divided into train/validation/test splits with a ratio of 8:1:1.

\paragraph{\textbf{Discussion}}
Our experiments with contrastive learning approach showed that the training process failed to converge. We attribute this failure to the query-dependent nature of code dependencies, which creates a contradictory optimization objective. During training, the same pair of code units may be treated as both similar and dissimilar depending on the query context, creating an ambiguous optimization landscape.
For instance, function \texttt{Func\_A} and function \texttt{Func\_B} may both be valid dependencies for a query, requiring their embeddings to be pulled closer. However, for a different query, \texttt{Func\_A} might be a true dependency, a positive sample, while \texttt{Func\_B} is a negative one, forcing the model to push their embedding apart. These contradictory signals result in conflicting gradient updates, destabilizing the training process. In contrast, the pairwise classification strategy circumvents this issue by evaluating each (\texttt{query}, \texttt{candidate}) pair independently.

For pairwise classification strategy, due to the natural class imbalance where negative dependencies vastly outnumber positive ones, to prevent model bias, we downsampled the negative class exclusively within the training set to match the number of positive samples, creating a balanced 1:1 ratio. The validation and test sets were kept in their original imbalanced state to ensure a realistic evaluation.

\subsection{\method Retriever - A Hybrid Approach}
\label{sec:hybrid}

\begin{figure*}
    \centering
    \includegraphics[width=\linewidth]{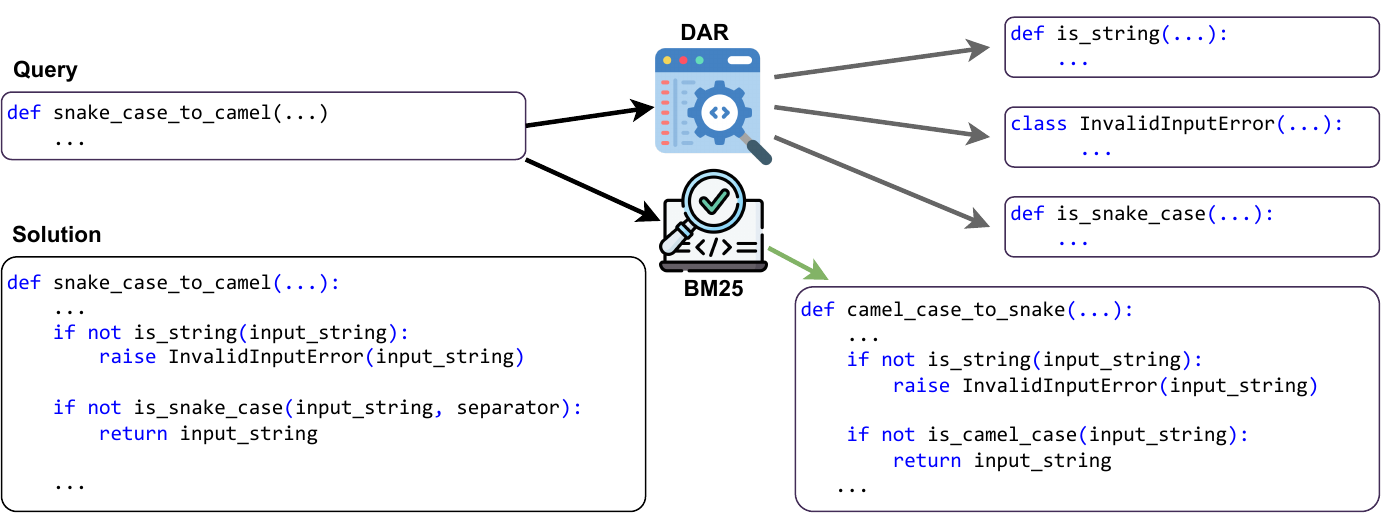}
    \caption{Example of retrieved results from DAR and BM25 for a query (incomplete function under generation) in the RepoExec dataset.}
    \label{fig:example_retrieval}
\end{figure*}

We hypothesize that dependency-aware retrieval provides the model with the essential ``\emph{materials}'' needed to implement the target function. However, this retriever might overlook the signal of \emph{how} these components should be used in practice (e.g., how to invoke a function or in which context a global variable is typically applied). In contrast, similarity-based retrieval can supply complementary information: by retrieving implementations that are lexically or semantically close to the target function, their bodies often illustrate concrete usage patterns of similar dependencies. Indeed, as illustrated in Figure~\ref{fig:example_retrieval}, when generating the \texttt{snake\_case\_to\_camel} function, DAR successfully identifies the correct dependencies that must be called. At the same time, BM25 can find the related function \texttt{camel\_case\_to\_snake}, which demonstrates how such dependencies are invoked in practice.

To combine these strengths, we design \textbf{\method Retriever}, a hybrid retrieval strategy that integrates dependency-aware retrieval (DAR) with similarity-based retrieval (i.e., BM25). By merging these complementary contexts, \method provides the model with both the necessary building blocks and usage examples, yielding a more complete and effective context for code generation.

\vspace{0.2cm}
\noindent \textbf{Inference process of \method:} During inference, the target function is passed through both the BM25 retriever and the DAR. 

\begin{itemize}[leftmargin=*]
    \item For BM25, we apply the standard formulation (Equation~\ref{eq:bm25}), following common practice by fixing $k_{1}=1.5$ and $b=0.75$. We then retrieve the top-5 most similar snippets. 

    \item For DAR, rather than using only the hard classification labels described in Section~\ref{sec:classification}, we leverage the prediction probability $p$ of class 1 to retain candidates, thereby reducing the risk of discarding true dependencies. To achieve this, we introduce a threshold hyperparameter $T$; a candidate dependency is retained if its predicted probability $p > T$. The threshold $T$ is tuned on the validation set following the strategy described below.

Our motivation in threshold selection is to capture as many ground-truth dependencies as possible (recall\(_1\)), while keeping the model's ability to detect false dependencies (recall\(_0\)) at an acceptable level. To balance this trade-off, we purpose the use of the Balanced Recall Penalty (BRP) score to select the optimal threshold. This metric's core function is to apply a penalty to the divergence between recall\(_1\) and recall\(_0\), scaled by a coefficient \(\alpha\) that reflects the class imbalance. The metric is defined as:
\begin{equation}
\text{BRP} = \text{recall}_1 - \alpha (\text{recall}_1 - \text{recall}_0)^2
\end{equation}
The penalty coefficient is calculated based on the class distribution as 
\begin{equation}
\alpha = \frac{1}{\lfloor \frac{N_{\text{class 1}} + N_{\text{class 0}}}{N_{\text{class 1}}} \rfloor}
\end{equation}
We performed a grid search using BRP on the validation set, which was constructed in Section \ref{sec:classification} with 46,141 negative instances (false dependencies) and 10,333 positive samples (true dependencies). Due to the class imbalance, the value of \(\alpha\) is thus determined to be 0.2.

\begin{table}[h!]
\centering
\caption{BRP scores at different threshold values $T$ on the validation set.}
\label{tab:brp_thresholds} 
\small{
\begin{tabular}{
    lcccccccc
}
\toprule
{\textbf{Threshold}} & 0.15 & 0.20 & \textbf{0.25} & 0.30 & 0.35 & 0.40 & 0.45 & 0.50 \\
\midrule
\textbf{BRP} & 0.8829 & 0.8852 & \textbf{0.8861} & 0.8838 & 0.8800 & 0.8755 & 0.8660 & 0.8522 \\
\bottomrule
\end{tabular}
}
\end{table}

As shown in Table \ref{tab:brp_thresholds}, a threshold of 0.25 yields the highest BRP score of 0.8861. Therefore, we select \(T = 0.25\) for our \method Retriever during inference.

\end{itemize}

%% file: section/4_experiment.tex
\section{Experiment Setup}

\subsection{Research Questions and Setup}

Our evaluation investigates the following research questions:

\vspace{0.2cm}
\noindent \textbf{RQ1:} \textit{How effective is Structure-Aware Indexing compared to Chunking-Based Indexing?}

$\hookrightarrow$ \textbf{Setup:} For chunking-based indexing, we follow \citet{zhang2023repocoderrepositorylevelcodecompletion} and segment each code file into fixed-size chunks of 2048 tokens with a 50\% overlap. In contrast, structure-aware indexing is performed as described in Section~\ref{sec:indexing}. For both indexing strategies, we employ the same BM25 retriever and Qwen2.5-Coder generator, and evaluate them on function-level code generation within a repository-level context.


\vspace{0.2cm}
\noindent \textbf{RQ2:} \textit{How do different retrieval approaches affect repository-level code generation performance?}

$\hookrightarrow$ \textbf{Setup:} In this RQ, we first evaluate the effectiveness of our proposed dependency-aware retriever (DAR) compared to prior similarity-based methods, including sparse retrieval with BM25 and dense retrieval with UniXCoder, in correctly retrieving the dependencies required to complete a function. For fairness, both strategies are applied over structure-aware indexing and evaluated on the test set described in Section~\ref{sec:training_data}, together with two considered benchmarks. We then extend the comparison to assess how these three retrieval methodologies, together with \method, affect performance on the repository-level code generation task.


\vspace{0.2cm}    
\noindent \textbf{RQ3:} \textit{How effective is \method compared to existing state-of-the-art approaches for repository-level code generation?}

$\hookrightarrow$ \textbf{Setup:} We compare \method against three state-of-the-art repository-level code generation approaches—RepoCoder \citep{zhang2023repocoderrepositorylevelcodecompletion}, RLCoder \citep{wang2024rlcoderreinforcementlearningrepositorylevel}, and Repoformer \citep{Repoformer} (Section~\ref{sec:related_work}); and further analyze their testing logs to provide a deeper assessment of output quality. In addition, we conduct an ablation study by removing functions, classes, and variables from Hydra’s search space to quantify the importance of granular components.


\vspace{0.2cm}
\noindent \textbf{RQ4:} \textit{How does the computational cost of running \method compare to state-of-the-art approaches for repository-level code generation?}

$\hookrightarrow$ \textbf{Setup:} We compare the retrieval time of \method against the baselines considered in RQ3 to demonstrate the practical efficiency of our approach. We report the mean, median, and maximum retrieval times across benchmarks to capture not only average efficiency but also the stability of each method under different query complexities.

\subsection{Benchmarks, Generator Backbones and Metrics}

\noindent \textbf{Benchmarks.}
In our evaluation, we utilize two recent benchmarks designed for repository-level code generation: 
\begin{itemize}
    \item \textbf{RepoExec}~\cite{nam2024repoexec}, a benchmark designed for the evaluation of repository-level code generation, with a focus on executability, functional correctness, and dependency utilization. The benchmark includes 355 problems. The core task challenges a model to generate code by effectively integrating provided code dependencies within a repository context. RepoExec provides an executable environment and a mechanism to automatically generate high-coverage test cases to robustly assess the functional correctness of the generated code.
    \item \textbf{DevEval}~\cite{DevEval}, a manually annotated benchmark created to closely align with real-world coding practices. DevEval comprises 1,825 test samples from 117 Python repositories across 10 popular domains. The core task is repository-level code generation, where a model is provided with a function signature, a detailed natural language requirement, and the full context of the repository.
\end{itemize}

\noindent \textbf{Generator Backbones.} For the generator models, we use the open-source Qwen2.5-Coder \citep{hui2024qwen25codertechnicalreport} with 1.5B and 7B parameters, as well as the closed-source GPT-4.1-mini \citep{openai2025gpt41}. This setup allows us to evaluate the adaptability of our method across different model types and scales. During generation, we employ nucleus sampling and generate 5 candidate outputs per query for all models.

\vspace{0.2cm}
\noindent \textbf{Metrics.} 
To evaluate the functional correctness of the generated code, we employ the widely-adopted execution-based metric, Pass@k~\cite{chen2021evaluatinglargelanguagemodels}, with k $\in \{1,3,5\}$. This metric assesses whether the generated code can successfully pass a set of predefined unit tests. In addition, we leverage the Dependency Invocation Rate (DIR) introduced by \citet{nam2024repoexec} to evaluate whether the generated code correctly calls the intended dependencies. This metric is computed as the size of the intersection between the set of dependencies invoked in the generated solution and those in the ground truth, divided by the total number of ground-truth dependencies.

%% file: section/5_evaluation.tex
\section{Evaluation}
\label{sec:evaluation}
In this section, we present the empirical results of our investigation, along with an in-depth analysis of our findings regarding the research questions.

\subsection*{RQ1: How effective is Structure-Aware Indexing compared to Chunking-Based Indexing?}
\label{RQ1}

\begin{figure*}
    \centering
    \includegraphics[width=\linewidth]{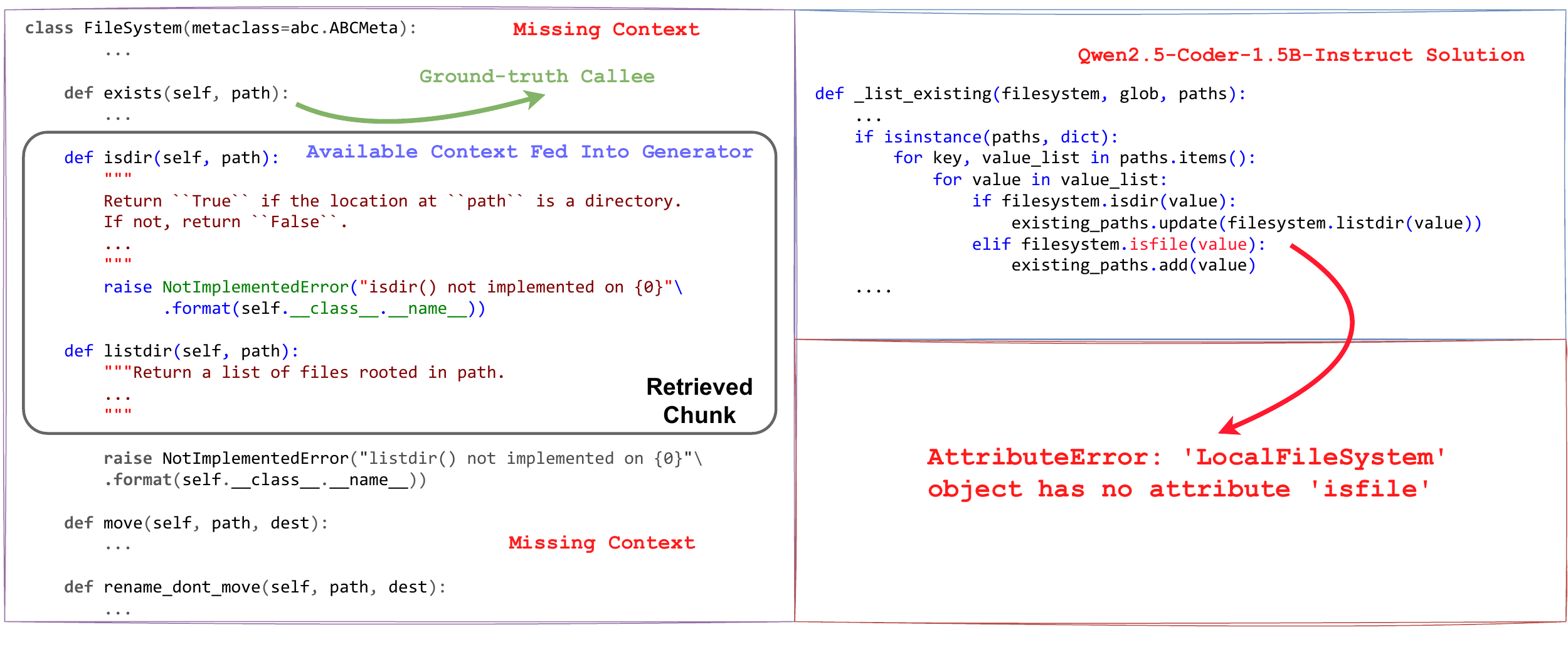}
    \caption{Example of contextual fragmentation under chunk-based retrieval. The retriever provides partial context (\texttt{isdir}, \texttt{listdir}) but misses the ground-truth dependency \texttt{exists}. As a result, the model hallucinates an \texttt{isfile} method, leading to an \texttt{AttributeError}.}
    \label{fig:RQ1}
\end{figure*}

\begin{table}[t]
\centering
\caption{Results of chunking-based vs. structure-aware context indexing on RepoExec and DevEval benchmarks with BM25 retriever, Qwen2.5-Coder-Instruct 1.5B and 7B versions as generator, reported in Pass@1/3/5.}
\label{tab:main_results_1}
\small{
\begin{tabular}{lllrrr}
\toprule
\textbf{Model Size} & \textbf{Benchmark} & \textbf{Indexing Strategy} & \textbf{Pass@1} & \textbf{Pass@3} & \textbf{Pass@5} \\
\midrule

\multirow{4}{*}{1.5B} 
& \multirow{2}{*}{RepoExec} 
& Chunking-based Indexing & 9.13 & 14.96 & 17.75 \\
& & \textbf{Structure-Aware Indexing} & \textbf{12.39} & \textbf{16.70} & \textbf{18.02} \\
\cmidrule(){2-6}

& \multirow{2}{*}{DevEval} 
& Chunking-based Indexing & 4.81 & 6.60 & 7.34 \\
& & \textbf{Structure-Aware Indexing} & \textbf{10.32} & \textbf{14.13} & \textbf{15.89} \\
\midrule

\multirow{4}{*}{7B} 
& \multirow{2}{*}{RepoExec} 
& Chunking-based Indexing & 14.03 & 19.32 & 21.69 \\
& & \textbf{Structure-Aware Indexing} & \textbf{20.79} & \textbf{27.61} & \textbf{30.14} \\
\cmidrule(){2-6}

& \multirow{2}{*}{DevEval} 
& Chunking-based Indexing & 8.46 & 10.88 & 11.78 \\
& & \textbf{Structure-Aware Indexing} & \textbf{16.67} & \textbf{21.73} & \textbf{23.78} \\

\bottomrule
\end{tabular}
}
\end{table}

To address this RQ, we evaluate repository-level code generation under two alternative context formulation strategies: chunking-base indexing and our proposed structure-aware context indexing. The results are presented in Table~\ref{tab:main_results_1}.

\paragraph{\textbf{Quantitative Findings.}}
The results demonstrate a consistent and substantial advantage for structure-aware context indexing. Across both model scales (1.5B and 7B) and benchmarks (RepoExec and DevEval), node-based context significantly outperforms chunk-based baselines. The performance gap is most pronounced on the challenging DevEval benchmark-for example, with QwenCoder-7B-Instruct, Pass@1 nearly doubles when using node extraction (16.67\% vs. 8.46\%). On RepoExec, we also observe steady improvements of 3–6\% in Pass@1, highlighting that this strategy is robust across repositories with diverse structures. These findings validate our hypothesis that preserving the syntactic and semantic integrity of code units yields a more coherent and complete view of the codebase, thereby improving the likelihood of correct generation.

\paragraph{\textbf{Qualitative Analysis: A Case Study on Contextual Fragmentation.}}
Quantitative metrics alone cannot fully explain why chunking degrades retrieval quality. To illustrate the failure modes, Figure~\ref{fig:RQ1} presents a case study on completing the \texttt{\_list\_existing} function, which relies on the \texttt{FileSystem} object. The chunk-based retriever captures fragments containing the \texttt{isdir} and \texttt{listdir} methods but fails to retrieve the critical \texttt{exists} method. Without this context, the model generates \texttt{filesystem.isfile(path)}, a plausible but hallucinated method inferred by analogy to \texttt{isdir}. The resulting \texttt{AttributeError}, illustrated in the traceback, is a direct consequence of semantic fragmentation: the retriever supplies incomplete and misleading context, which forces the model into an erroneous inference.

\begin{RQsummary}
\textbf{[RQ1] Summary:} Unlike natural language, code cannot be reliably segmented into arbitrary chunks without breaking logical boundaries. Chunk-based indexing often fragments dependencies, yielding incomplete or misleading context that leads models to hallucinate calls and produce incorrect code. Structure-aware context indexing instead preserves functions, classes, and variables as coherent units, providing a more faithful representation of the repository. This approach significantly improves retrieval quality and code generation, surpassing chunking-based strategies by a wide margin, with around $2\times$ gains on DevEval.
\end{RQsummary}

\subsection*{RQ2: How do different retrieval approaches affect repository-level code generation performance?}

\begin{table*}[th]
\centering
\caption{Comparison of retrieval methods. The top section shows overall performance in terms of Recall, Precision, and F1. The bottom section provides a fine-grained breakdown of Recall by different dependency types, functions (FRecall), classes (CRecall), and variables (VRecall)}
\label{tab:dependency_retrieval}
\begin{tabular}{lcccccc}
\toprule
\multirow{2}{*}{\textbf{Retriever}} & \multicolumn{3}{c}{\textbf{RepoExec}} & \multicolumn{3}{c}{\textbf{DevEval}} \\
\cmidrule(lr){2-4} \cmidrule(lr){5-7}
 & Recall & Precision & F1 & Recall & Precision & F1 \\
\midrule
Sparse Retriever & 0.5378 & 0.0809 & 0.1406 & 0.5734 & 0.1243 & 0.2044 \\
Dense Retriever & 0.5483 & 0.0824 & 0.1432 & 0.5183 & 0.1124 & 0.1848 \\
Dependency-aware Retriever & \textbf{0.9223} & \textbf{0.1086} & \textbf{0.1957} & \textbf{0.8884} & \textbf{0.1559} & \textbf{0.2652} \\
\midrule
 & FRecall & CRecall & VRecall & FRecall & CRecall & VRecall \\
\midrule
Sparse Retriever & 0.6522 & 0.6667 & 0.3174 & 0.7216 & 0.6562 & 0.3227 \\
Dense Retriever & 0.6957 & 0.5196 & 0.3832 & 0.7557 & 0.5191 & 0.3506 \\
Dependency-aware Retriever & \textbf{0.9420} & \textbf{0.8922} & \textbf{0.9162} & \textbf{0.8718} & \textbf{0.9194} & \textbf{0.8812} \\
\bottomrule
\end{tabular}

\end{table*}

\begin{table*}[t]
\centering
\caption{Code generation results with different retrieval strategies on RepoExec and DevEval benchmarks using Qwen2.5-Coder models (1.5B and 7B) as generator. Results are reported in Pass@1/3/5 and Dependency Invocation Rate (DIR).}
\label{tab:main_results_4}
\begin{adjustbox}{width=\textwidth}

\begin{tabular}{llccccccc}
\toprule
\multirow{2}{*}{\textbf{Model Size}} & \multirow{2}{*}{\textbf{Retriever}} & 
\multicolumn{4}{c}{\textbf{RepoExec}} & \multicolumn{3}{c}{\textbf{DevEval}} \\
\cmidrule(lr){3-6} \cmidrule(lr){7-9}
& & Pass@1 & Pass@3 & Pass@5 & DIR & Pass@1 & Pass@3 & Pass@5 \\
\midrule
\multirow{4}{*}{1.5B} & Sparse Retriever & {12.39} & {16.70} & {18.02} & {39.93} & {10.32} & {14.13} & {15.89} \\
& Dense Retriever & 12.73 & 19.04 & 21.40 & 39.86 & 10.24 & 13.89 & 15.56\\
& Dependency-aware Retriever & 13.46 & 18.02 & 20.28 & 43.81 & 10.43 & \textbf{14.58} & \textbf{16.22}\\
& \method Retriever& \multicolumn{1}{c}{\bfseries 15.72} & \multicolumn{1}{c}{\textbf{21.30}} & \multicolumn{1}{c}{\textbf{23.38}} & {\textbf{44.61}} & \multicolumn{1}{c}{\textbf{10.71}} & \multicolumn{1}{c}{14.50} & \multicolumn{1}{c}{16.05} \\

\midrule
\multirow{4}{*}{7B} & Sparse Retriever & {20.79} & {27.61} & {30.14} & {49.67} & {16.67} & {21.73} & {23.78} \\
& Dense Retriever & {19.89} & {26.45} & {29.58} & {46.38} & {15.93} & {20.92} & {22.79} \\
& Dependency-aware Retriever & {21.97} & {29.89} & {32.96} & {52.24} & {16.84} &{21.93} & {23.95} \\
& \method Retriever& \multicolumn{1}{c}{\bfseries 23.32} & \multicolumn{1}{c}{\textbf{31.32}} & \multicolumn{1}{c}{\textbf{34.36}} & {\textbf{53.46}} & \multicolumn{1}{c}{\textbf{17.27}} & \multicolumn{1}{c}{\textbf{22.44}} & \multicolumn{1}{c}{\textbf{24.44}} \\

\bottomrule
\end{tabular}
\end{adjustbox}
\end{table*}

First, we evaluate the effectiveness of the proposed Dependency-Aware Retrieval (DAR) in comparison to sparse and dense retrieval methods for acquiring the dependencies required to implement a target function. Experiments are conducted on our constructed test dataset (Sections~\ref{sec:training_data} and \ref{sec:classification}) as well as on the RepoExec and DevEval benchmarks. We report Precision, Recall, and F1 for true dependency class. Since our primary objective is to ensure that true dependencies are not missed, Recall on the positive class (relevant dependencies) serves as the main evaluation metric. At the same time, retrieving irrelevant dependencies introduces noise, which reflects the model’s ability to correctly filter candidates. Although such noise should be mitigated, it also provides insight into how effectively the retriever selects useful materials for the generator. 

Table~\ref{tab:dependency_retrieval} reports the performance of sparse, dense, and dependency-aware retrieval (DAR) in detecting true dependencies. On RepoExec, both sparse and dense retrievers achieve moderate recall (around 0.53–0.55), but their very low precision (<0.09) shows that many retrieved candidates are spurious. In contrast, DAR reaches a recall of 0.92, substantially higher than either baseline, confirming its ability to capture the majority of ground-truth dependencies. Although precision remains modest, this trade-off is acceptable: for code generation, failing to retrieve true dependencies is far more damaging than including some irrelevant context. On DevEval, the results are consistent. Sparse and dense methods again show moderate recall (0.51–0.57) and low precision (0.11–0.12), while DAR achieves a recall of 0.89 with the highest F1 among all methods. These gains underscore the robustness of DAR across different benchmarks. Overall, the results validate our design choice to prioritize recall of true dependencies, ensuring that the generator is supplied with the essential building blocks for producing correct and executable code, even if some noise is introduced.

\begin{figure}[t]
\centering
\small
\centering
\text{(a) Target Function Prompt}
\begin{lstlisting}[numbers=none]
def is_url(input_string: Any, allowed_schemes: Optional[List[str]] = None) -> bool:
    """
    Check if a string is a valid URL.
    [...]
    """
\end{lstlisting}
\hdashrule{\linewidth}{0.5pt}{3pt 3pt}
\centering
\text{(b) Solution generated with \method}
\begin{lstlisting}[numbers=none]
def is_url(input_string: Any, allowed_schemes: Optional[List[str]] = None) -> bool:
    [...]
    if not is_full_string(input_string):
        return False
    [...]
\end{lstlisting}
\hdashrule{\linewidth}{0.5pt}{3pt 3pt}
\centering
\text{(c) Solution generated with BM25}
\begin{lstlisting}[numbers=none]
def is_url(input_string: Any, allowed_schemes: Optional[List[str]] = None) -> bool:
    [...]
    if not isinstance(input_string, str):
        raise InvalidInputError(input_string)
    [...]
\end{lstlisting}
\vspace{-11pt}
\caption{Case study comparing the generated code by \method and text-based retrieval on \texttt{task\_id} 18 (RepoExec).}
\label{fig:case_study}
\end{figure}

We further conduct a fine-grained analysis of retrieval performance across different dependency types, namely functions, classes, and variables. As shown in Table~\ref{tab:dependency_retrieval}, similarity-based retrievers achieve their highest recall on function dependencies (0.65–0.75). This is understandable since the task is framed as function-level completion and many pretrained models are optimized using function-level code snippets \citep{husain2019codesearchnet, nguyen2023vault, guo2022unixcoderunifiedcrossmodalpretraining, wang2021codet5}. These retrievers struggle more with class dependencies, though their recall remains at a moderate level. In contrast, variables remain the most challenging to retrieve for similarity-based approaches, with recall below $0.4$. To our knowledge, variables have received little explicit consideration in prior work, either during pretraining or fine-tuning of retrieval and code generation models. Nevertheless, variables-particularly constants and configuration values-are frequently leveraged in implementing target functions. Missing these dependencies not only reduces generation accuracy but also risks overlooking the underlying human intent of the code. By comparison, the proposed DAR achieves consistently high recall across all dependency types, reaching 0.94/0.89/0.92 on RepoExec and 0.87/0.92/0.88 on DevEval for functions, classes, and variables, respectively. This balanced coverage highlights the robustness of DAR: It not only significantly improves performance on function type, where similarity-based methods already perform best, but also substantially improves retrieval for classes and variables, which are critical for repository-level code generation but often missed by similarity-driven approaches.

Additionally, we evaluate our method within the full RAG pipeline for repository-level code generation, in order to facilitate a comprehensive assessment of its effectiveness. Table~\ref{tab:main_results_4} reports code generation results on RepoExec and DevEval using different retrieval strategies with Qwen2.5-Coder models (1.5B and 7B). We can observe that DAR achieves consistently higher Pass@k scores than similarity-based retrievers, indicating that providing true dependencies offers greater value to the generator than merely retrieving lexically or semantically similar snippets. Besides, it also achieves higher DIR compared to sparse and dense methods (e.g., 43.81 vs. 39.9 at 1.5B, and 52.24 vs. 46–49 at 7B), confirming its superior ability to recover true dependencies, thereby reducing the risk of code smells and technical debt. However, the performance gap remains modest, suggesting that similarity-based context still provides complementary value. Indeed, the hybrid \method Retriever consistently delivers the best overall performance. At the 1.5B scale, \method raises RepoExec Pass@1 to 15.72, surpassing sparse (12.39), dense (12.73), and DAR (13.46) by more than 2\%. Similar improvements are observed in Pass@3, Pass@5, and DIR metrics. On the more challenging DevEval benchmark, \method achieves Pass@1 scores of 10.71 and 17.27 with the 1.5B and 7B models, respectively, outperforming all baselines. These results demonstrate that similarity-based context complements dependency context by providing signals on \emph{how} dependencies are used, thereby helping the generator produce code that is both functionally correct and aligned with the current repository state, further strengthening our hypothesis and discussion in Section~\ref{sec:hybrid}.

\vspace{0.2cm}
\noindent \textbf{Case Study:} We further demonstrate the robustness of Hydra compared to text-based retriever through a representative case study, illustrated in Figure~\ref{fig:case_study}. The task (RepoExec, \texttt{task\_id} 18) requires completing the function \texttt{is\_url}, whose goal is to validate whether an input is a valid URL, as specified by its signature and docstring. Correctly fulfilling this specification depends on checking whether the input is a valid string, which is implemented in the auxiliary function \texttt{is\_full\_string}. As shown in the figure, \method, via DAR, successfully retrieves the dependent function \texttt{is\_full\_string}, enabling the LLM to reuse the existing API and generate a correct implementation. In contrast, similarity-based retrievers fail to retrieve this dependency, causing the LLM to re-implement the validation logic on its own. Without access to the required dependent component, the generated code deviates from the intended behavior and frequently leads to incorrect implementations. This case study highlights \method’s advantage in retrieving critical dependencies that text-based retrievers often overlook.


\begin{RQsummary}
\textbf{[RQ2] Summary:} Similarity-based retrieval methods, such as sparse retrieval or dense retrieval with a general encoder, frequently overlook a substantial portion of required dependencies-particularly variables and classes. In contrast, the proposed Dependency-Aware Retrieval (DAR) demonstrates greater robustness in accurately identifying candidate dependencies. 

Furthermore, DAR consistently achieves higher Pass@k and DIR than similarity-based retrievers, showing that true dependencies provide greater value to the generator and help reduce risks such as code smells and technical debt. Nonetheless, similarity-based context still adds complementary usage signals, and the hybrid \method Retriever leverages both to achieve the strongest results.
\end{RQsummary}

\subsection*{RQ3: How effective is \method compared to existing state-of-the-art approaches for repository-level
code generation?}

\paragraph{\textbf{\method vs. Baselines.}} Table~\ref{tab:main_results} compares \method with several representative baselines, including RepoCoder, RepoFormer, and RLCoder, as well as no-context generation. Across all model families and sizes, \method consistently delivers the best performance, establishing itself as a new state of the art in repository-level code generation. Notably, while existing methods already improve over the no-context baseline by supplying additional cross-file information, their reliance on chunking or similarity-based retrieval limits the completeness and utility of the retrieved context. In contrast, \method’s hybrid strategy-anchoring generation on true dependencies, augmenting them with similarity-based usage signals, and leveraging structure-aware indexing-supplies the model with both the essential building blocks and clear guidance on how to apply them effectively. This advantage holds for both open-source and closed-source models, and across benchmarks, underscoring the robustness and generalizability of our approach. 

A particularly striking result is that, on both benchmarks, the 1.5B model equipped with our \method retriever surpasses much larger baselines: it outperforms RepoCoder on RepoExec and even exceeds RepoFormer on DevEval with a 7B model (15.72 vs. 14.82 and 10.71 vs. 10.41 in Pass@1, respectively). This demonstrates that high-quality retrieval can narrow or even bridge the performance gap between small and large models, highlighting the potential value of our approach in resource-constrained settings.

\begin{table*}[thb]
\centering
\caption{Comparison of \method with prior retrieval-based approaches and no-context baselines. Results are reported in Pass@1/3/5 using close sourced GPT-4.1 mini and open source Qwen2.5-Coder-1.5B and 7B Instruct models.}
\label{tab:main_results}
\small{
\begin{tabular}{llcccccc}
\toprule
\multirow{2}{*}{\textbf{Generator}} & \multirow{2}{*}{\textbf{Method}} & 
\multicolumn{3}{c}{\textbf{RepoExec}} & \multicolumn{3}{c}{\textbf{DevEval}} \\
\cmidrule(lr){3-5} \cmidrule(lr){6-8}
& & Pass@1 & Pass@3 & Pass@5 & Pass@1 & Pass@3 & Pass@5 \\
\midrule
\multirow{5}{*}{GPT-4.1 mini} & No Context & {21.58} & {24.42}& {25.63} & {19.72} & {23.19} & {24.71}\\
& RepoCoder & {22.20} & {26.08} & {27.89} & {17.48} & {23.15} & {25.70}\\
& RepoFormer & {39.15} & {42.42} & {43.94} & {30.89} & {34.21} & {35.40}\\
& RLCoder & {38.14} & {42.17} & {43.38} & {29.46} & {32.76} & {34.14} \\
& \textbf{\method} & \multicolumn{1}{c}{\bfseries 43.55} & \multicolumn{1}{c}{\bfseries 45.72} & \multicolumn{1}{c}{\bfseries 46.48} & \multicolumn{1}{c}{\bfseries 31.91} & \multicolumn{1}{c}{\bfseries 35.56} & \multicolumn{1}{c}{\bfseries 36.99 } \\

\midrule
\multirow{5}{*}{QwenCoder-1.5B-Instruct} & No Context & {5.75} & {8.31} & {9.30} & {3.53} & {5.20} & {5.97}\\
& RepoCoder & {7.15} & {11.72}  & {14.37} & {4.54} & {8.08} & {9.81} \\
& RepoFormer & {11.15} & {16.42} & {18.87} & {5.58} & {7.94} & {8.99} \\
& RLCoder & {14.87} & {21.04} & \bfseries {23.94} & {9.34} & {12.90} & {14.47} \\
& \textbf{\method} & \multicolumn{1}{c}{\bfseries 15.72} & \multicolumn{1}{c}{\bfseries 21.30} & \multicolumn{1}{c}{23.38} & \multicolumn{1}{c}{\bfseries 10.71} & \multicolumn{1}{c}{\bfseries 14.50} & \multicolumn{1}{c}{\bfseries 16.05} \\
\midrule
\multirow{5}{*}{QwenCoder-7B-Instruct} & No Context & {13.30} & {17.04} & {18.03} & {7.10} & {9.16} & {10.03}\\
& RepoCoder & {14.82} & {21.99} & {25.07} & {6.39} & {10.63} & {12.82} \\
& RepoFormer & {17.69} & {25.04} & {28.45} & {10.41} & {13.68} & {14.90} \\
& RLCoder & {20.17} & {23.69} & {27.61} & {13.00} & {17.67} & {19.61} \\
& \textbf{\method} & \multicolumn{1}{c}{\bfseries 23.32} & \multicolumn{1}{c}{\bfseries 31.32} & \multicolumn{1}{c}{\bfseries 34.36} & \multicolumn{1}{c}{\bfseries 17.27} & \multicolumn{1}{c}{\bfseries 22.44} & \multicolumn{1}{c}{\bfseries 24.44} \\
\bottomrule
\end{tabular}
}
\end{table*}

\begin{figure}[t] 
    \centering
    \includegraphics[width=\textwidth
    ]{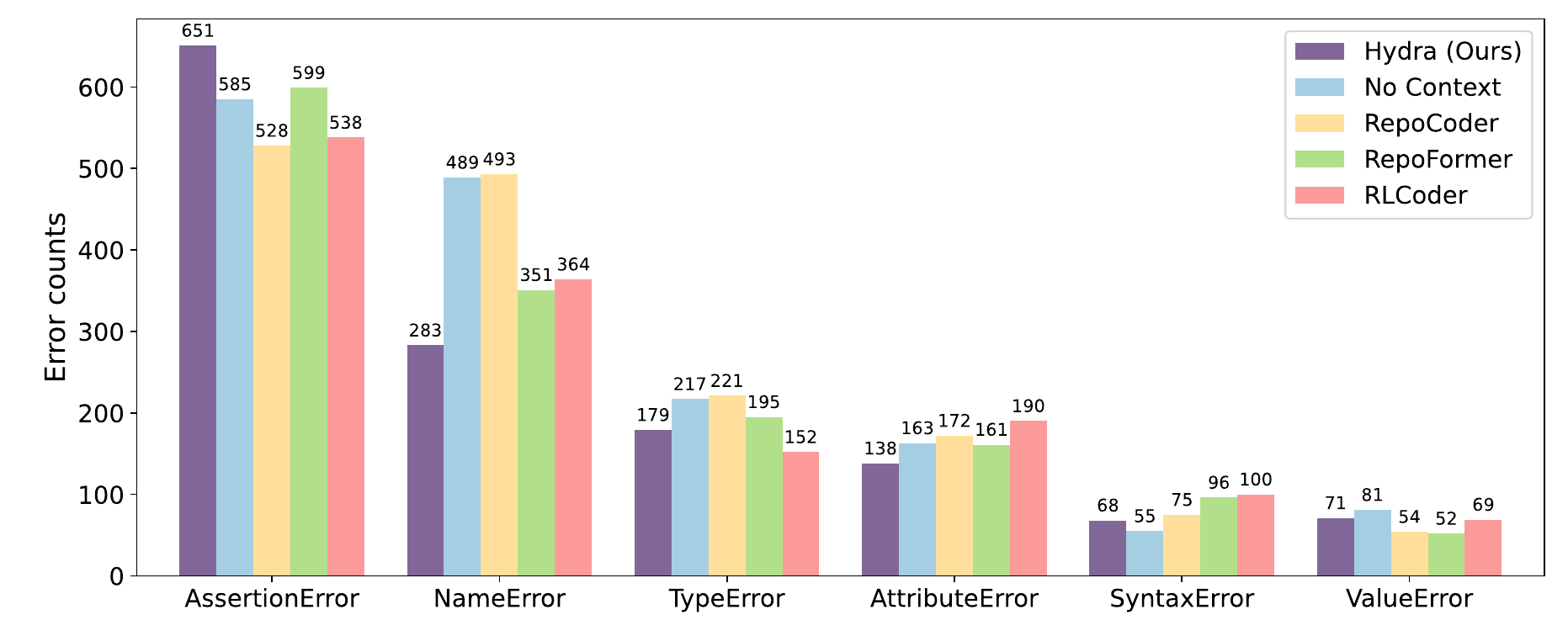} 
    \caption{Distribution of error types in generated code for different repository-level code generation methods, evaluated with Qwen2.5-Coder-1.5B-Instruct on RepoExec.}
    \label{fig:error-distribution}
\end{figure}

\paragraph{\textbf{Failure Modes of Generated Code.}} We further investigate the failure modes of generated code and present the error statistics in Figure \ref{fig:error-distribution}. Compared to baselines (No Context, RepoCoder, RepoFormer, and RLCoder), \method significantly reduces critical errors such as NameError, TypeError, and AttributeError. These error categories are particularly indicative of model hallucination, where the generator fabricates dependencies that do not exist in the repository. The substantial reduction demonstrates the effectiveness of \method in preserving code structure, retrieving true dependencies, and providing accurate usage guidance, thereby helping the model avoid spurious or inconsistent component calls.

At the same time, we observe a relative increase in AssertionError. Rather than signaling degraded generation quality, this rise stems from richer test execution coverage: with correct dependencies in place, the generated functions are more likely to run and reach assertion checks, exposing deeper logical mismatches rather than failing prematurely due to missing symbols. This trade-off suggests that \method shifts errors from structural failures toward more nuanced behavioral validation, reflecting an important step toward generating executable and repository-consistent code.

\begin{table*}[t]
\centering
\caption{Ablation study evaluating the contribution of different granular retrieval components in Hydra, including function-level, class-level, and variable-level retrieval.}
\label{tab:ablation_results}
\small{
\begin{tabular}{llcccccc}
\toprule
\multirow{2}{*}{\textbf{Generator}} & \multirow{2}{*}{\textbf{Method}} & 
\multicolumn{3}{c}{\textbf{RepoExec}} & 
\multicolumn{3}{c}{\textbf{DevEval}} \\
\cmidrule(lr){3-5} \cmidrule(lr){6-8}
& & Pass@1 & Pass@3 & Pass@5 & Pass@1 & Pass@3 & Pass@5 \\
\midrule

\multirow{4}{*}{QwenCoder-1.5B-Instruct}
& \textbf{\method}
& \bfseries 15.72 & \bfseries 21.30 & \bfseries 23.38
& \bfseries 10.71 & \bfseries 14.50 & \bfseries 16.05 \\

& \textit{w/o-Function}
& 13.12 & 15.69 & 16.62
& 8.58 & 12.43 & 14.30 \\

& \textit{w/o-Class}
& 11.94 & 17.18 & 19.15 
& 8.24 & 12.03 & 14.02 \\

& \textit{w/o-Variable}
& 13.01 & 17.86 & 20.00
& 9.21 & 13.35 & 15.29 \\

\midrule

\multirow{4}{*}{QwenCoder-7B-Instruct}
& \textbf{\method}
& \bfseries 23.32 & \bfseries 31.32 & \bfseries 34.36
& \bfseries 17.27 & \bfseries 22.44 & \bfseries 24.44 \\

& \textit{w/o-Function}
& 17.97 & 24.54 & 27.32
& 14.76 & 19.58 & 21.48 \\

& \textit{w/o-Class}
& 20.73 & 27.15 & 30.14
& 14.15 & 19.04 & 21.21 \\

& \textit{w/o-Variable}
& 21.01 & 28.17 & 30.42
& 15.72 & 20.07 & 21.81 \\

\bottomrule
\end{tabular}
}
\end{table*}

\paragraph{\textbf{Ablation study.}} To comprehend the individual contributions of each component type to Hydra's effectiveness, we conduct an ablation study by systematically removing function-level, class-level, and variable-level retrieval from the search space. Table~\ref{tab:ablation_results} presents the results across both benchmarks using Qwen2.5-Coder-1.5B-Instruct and Qwen2.5-Coder-7B-Instruct. Here, \textit{``w/o-Function''}, \textit{``w/o-Class''}, and \textit{``w/o-Variable''} denote variants where function-level, class-level, and variable-level components are respectively removed from the search space of \method. From Table~\ref{tab:ablation_results}, we observe that removing any retrieval granularity consistently degrades performance across both benchmarks and model scales. On RepoExec and DevEval, ablating function-level and class-level retrieval leads to a pronounced performance drop, with Pass@1 decreasing by up to approximately 4\%. In contrast, ablating variable-level retrieval results in a comparatively smaller degradation, with Pass@1 typically dropping by around 1-2\%. These results indicate that function- and class-level retrieval provide more essential API and structural signals, while variable-level retrieval plays a complementary role.

\begin{RQsummary}
\textbf{[RQ3] Summary:} \method establishes a new state of the art in repository-level code generation, consistently surpassing baselines such as RepoCoder, RepoFormer, and RLCoder across models and benchmarks. Its hybrid design, combining dependency-aware retrieval, similarity-based usage signals, and structure-aware indexing-allows smaller models to surpass larger ones, highlighting the value of high-quality retrieval. Error analysis further shows that \method reduces hallucination-related failures while shifting errors toward deeper behavioral checks, leading to more executable and repository-consistent code. Additionally, The ablation study shows that Hydra benefits from all granular retrieval components, with function- and class-level retrieval contributing most, while variable-level retrieval provides complementary gains that further improve generation quality.
\end{RQsummary}

\subsection*{\textbf{RQ4:} How does the computational cost of running \method compare to state-of-the-art approaches for repository-level code generation?}

In this RQ, we evaluate the retrieval latency of different methods to assess their practicality for repository-level code generation. Table~\ref{tab:time_latency} compares the retrieval latency of different repository-level code generation methods across benchmarks. The results highlight clear differences in efficiency stemming from how each method structures its retrieval process.

RepoCoder exhibits by far the highest latency, with maximum times extending to several tens of seconds. This inefficiency stems from its iterative RAG design, where the retriever repeatedly queries and refines the context through multiple interactions with the codebase, leading to high computational cost. In contrast, RLCoder achieves relatively reasonable efficiency, with most queries completing under one second. However, it still suffers from outliers, reflected in large gaps between mean and median latencies, suggesting instability in retrieval time. RepoFormer demonstrates the fastest median retrieval time across both benchmarks, benefiting from its selective retrieval strategy, which invokes RAG only when the model’s generation is predicted to benefit from additional context. However, this comes at the cost of occasionally very high maximum latencies, reflecting the overhead when retrieval is triggered.

By comparison, our proposed \method achieves consistently low latency while maintaining stable performance across queries, owing to its context-aware retrieval design. Instead of exhaustively scanning the entire repository, \method anchors retrieval on call-graph dependencies at the function and file levels, restricting the search space to the current file and its imported modules (via import connections in Figure~\ref{fig:indexing}). This is particularly effective in practice, as many target functions reside in leaf files that do not import other modules, resulting in a very small candidate pool or even no candidate dependencies. In contrast, prior works traverse the entire codebase without considering dependency structure, resulting in unnecessary computation over large amounts of irrelevant code, higher average latency, and frequent extreme outliers. Moreover, \method incorporates a lightweight binary classification step to predict whether a candidate dependency is relevant to the target function, further narrowing the retrieved context and reducing redundant processing. From a complexity perspective, \method incurs a retrieval cost of $O(c)$, where $c$ corresponds to the size of the current file and its imported dependencies, whereas existing approaches exhibit an $O(n)$ cost with respect to the full repository size. Since $n$ grows much faster than $c$ in real-world codebases, this gap becomes increasingly pronounced as repositories scale. Consequently, \method delivers substantial retrieval speedups without sacrificing generation quality, making it well-suited for practical repository-level code generation scenarios where both accuracy and low latency are critical.


\begin{table*}[t]
\centering

\caption{Latency comparison (in milliseconds per example) of different repository-level code generation methods. Results are reported in terms of minimum, maximum, mean, and median latency.}
\label{tab:time_latency}
\small{
\begin{tabular}{lcccccccc}
\toprule
\multirow{2}{*}{\textbf{Method}} & \multicolumn{4}{c}{\textbf{RepoExec}} & \multicolumn{4}{c}{\textbf{DevEval}} \\
\cmidrule(lr){2-5} \cmidrule(lr){6-9}
& Min & Max & Mean & Median & Min & Max & Mean & Median \\
\midrule
{RLCoder} & {91.71} & {2927.08} & {644.92} & {347.16} & {91.49} & \textbf{1824.29} & {894.40} & {821.43}\\

{RepoFormer} & {13.59} & {27286.85} & {1766.28} & \textbf{18.44} & {12.40} & {34336.94} & {4432.41} & \textbf{18.86}\\

{RepoCoder} & {919.64} & {34072.90} & {7212.16} & {5201.46} & {566.52} & {24306.62} & {5096.50} & {4618.88}\\

{\textbf{\method}} & \textbf{0.01} & \textbf{1836.32} & \textbf{362.95} & {277.80} & \textbf{0.01} & {21234.01} & \textbf{432.86} & {221.22}\\
\bottomrule
\end{tabular}
}
\end{table*}

\begin{RQsummary}
\textbf{[RQ4] Summary:} \method delivers both state-of-the-art accuracy and the lowest, most stable latency by anchoring retrieval on dependencies and narrowing the search space to relevant files.
\end{RQsummary}




%% file: section/7.Threads.tex
\section{Threats to Validity}



Our study, while demonstrating the effectiveness of \method, is subject to several potential threats to internal and external validity.

\subsection{Internal Validity} 

\noindent \textbf{Retriever Backbone Choice:} A potential threat lies in the choice of backbone model for the retriever. In this work, we only evaluate UniXCoder as the encoder for dependency detection. While UniXCoder provides a strong baseline and stable convergence in our setting, alternative backbones (e.g., CodeBERT \citep{feng2020codebert}, GraphCodeBERT \citep{guographcodebert}, or newer code-specific encoders) might lead to different retrieval effectiveness. As a result, our conclusions about retriever performance may be partly tied to this specific backbone choice.

\vspace{0.2cm}
\noindent \textbf{Data Contamination: } For the Hydra retriever,  we filter out any repositories that overlap with the evaluation datasets during training data construction, preventing direct data leakage. Besides, we computed the entropy of DAR’s predictions on RepoExec and DevEval. The relatively high average entropy (0.3634 on RepoExec and 0.4035 on DevEval, with a maximum of 0.6) suggests that DAR is not biased toward memorized patterns and is minimally affected by potential data contamination. Regarding the risk of generator (LLMs) being pretrained on the evaluation data, we empirically show that plain LLMs without Hydra’s retrieval module perform substantially worse than Hydra across all settings. In particular, removing retrieval leads to performance drops of over 10\% on the QwenCoder variants and over 20\% on GPT-4.1-mini. These results indicate that Hydra’s performance gains primarily stem from its retrieval mechanism rather than memorization, mitigating the potential data contamination.

\subsection{External Validity}

\noindent \textbf{Scope Limited to Python Benchmarks:} Our experiments are restricted to Python and two benchmarks, RepoExec and DevEval. While these benchmarks represent realistic and challenging settings, results may not directly generalize to other programming languages, ecosystems, or repository structures. The focus on function-level generation, while aligned with benchmark design, also narrows the scope, leaving the effectiveness of \method for higher-level tasks (e.g., class or module generation) as future work.

\vspace{0.2cm}
\noindent \textbf{Limited choice of LLMs:}
Our experiments involve two types of LLMs as the generator: Qwen2.5-Coder and GPT-4.1 mini. Although our method is model-agnostic, evaluation on a broader range of models lies beyond the scope of this work. We believe, however, that our selected models are representative, covering both state-of-the-art open-source and closed-source LLMs for code generation.

%% file: section/8_conclusion.tex
\section{Conclusion}

In this work, we introduced \method, a framework for repository-level code generation that leverages structure-aware indexing, dependency-aware retrieval, and a hybrid strategy combining dependencies with similarity-based signals. Experiments on two challenging benchmarks, RepoExec and DevEval, show that \method consistently surpasses baselines, establishes a new state of the art, and even enables small models to rival larger ones. Ablation studies further confirm the effectiveness of incorporating granular context at multiple levels, including functions, classes, and variables, which contributes substantially to performance gains. Error analysis reveals that \method reduces hallucination-related failures and shifts errors toward deeper behavioral validation, resulting in more executable and repository-consistent code. Moreover, our time complexity and latency evaluation demonstrate that \method achieves practical efficiency by restricting retrieval to dependency-localized context rather than repository-wide scanning. Overall, these findings highlight the importance of moving beyond NLP-inspired chunking and similarity retrieval toward dependency-centered approaches for practical and reliable repository-level code generation.

%% file: section/9.data-avai.tex
\section{Data Availability}
A replication package that contains the source code and datasets for this paper is available at \url{https://github.com/solis-team/Hydra}.